\documentclass[12pt,preprint]{aastex}

\usepackage{mathptmx}
\usepackage{courier}
\usepackage{graphicx}
\usepackage{colordvi}
\usepackage{natbib}
\usepackage{color}

\newcommand{\FeI}{\ion{Fe}{1}}

\newcommand{\CHAX}{CH~A$^{2}\Delta$--X$^{2}\Pi$}

\shortauthors{Uitenbroek et al.}
\shorttitle{The contrast in the G-Band}

\begin{document}

%
%

\title{The discrepancy in G-band contrast: Where is the quiet Sun?}

\author{H.~Uitenbroek, A.~Tritschler and T.~Rimmele}
\affil{National Solar Observatory/Sacramento Peak\footnote{Operated by the %
       Association of Universities for Research in Astronomy, Inc. (AURA), %
       for the National Science Foundation}, P.O.~Box 62, Sunspot,
       NM-88349, U.S.A.}
\email{huitenbroek@nso.edu, ali@nso.edu, trimmele@nso.edu}

\date{Version \today}

%
%

\begin{abstract}
We compare the rms contrast in observed speckle reconstructed
G-band images with synthetic filtergrams computed from two
magneto-hydrodynamic simulation snapshots.
The observations consist of 103 bursts of 80 frames each taken at the
Dunn Solar Telescope (DST), sampled at twice the diffraction limit of the
telescope.
The speckle reconstructions account for the performance of the Adaptive
Optics (AO) system at the DST to supply reliable photometry.
We find a considerable discrepancy in the observed rms contrast
of 14.1\% for the best reconstructed images, and the synthetic
rms contrast of 21.5\% in a simulation snapshot thought to be
representative of the quiet Sun.
The areas of features in the synthetic filtergrams that have positive
or negative contrast beyond the minimum and maximum values in the
reconstructed images have spatial scales that should be resolved.
This leads us to conclude that there are fundamental differences
in the rms G-band contrast between observed and computed filtergrams.
On the basis of the substantially reduced granular contrast of 16.3\%
in the synthetic plage filtergram we speculate that the quiet-Sun may contain
more weak magnetic field than previously thought.
\end{abstract}

\keywords{Sun: granulation --- Sun: photosphere ---
          techniques: high angular resolution --- (magnetohydrodynamics:) MHD}

%
%

\section{Introduction}\label{sec:introduction}
Magneto-hydrodynamic simulations are increasingly successful in
reproducing the observable effects of magneto-convection on the solar surface.
In achieving this goal they provide not only insight into the physics
that drives convective patterns at the surface, but also into the
dynamics of the solar interior, where at best only helioseismological
observations can probe.
Although the simulations are typically claimed to be parameter free,
and therefore cannot be adjusted at will to match observations,
they include arbitrary elements in the choice of the physical approximations
that are implemented to keep the numerical problem tractable.
Significant limitations are restricting radiative transfer to a small
number of representative frequency bins and discrete directions,
and strongly reducing the Reynolds numbers in the simulation to
values that are orders of magnitude less than in reality
      \citep{Stein+Nordlund1998,Voegler_etal2005,Schaffenberger_etal2006}.
The latter restriction limits the smallest spatial scales at which
structures can occur and how the dynamics on these scales can possibly
feed back to larger scales.

One of the major successes of solar convection simulations is the
very accurate reproduction of spatially and temporally averaged shapes
of weak photospheric lines
     \citep{Asplund+others2000,Stein+Nordlund2000}.
These line shapes, characterized by an asymmetry towards the blue,
are the intricate product of correlation between up- and downflows,
and their respective brightness differences and area asymmetries.
Because the comparison is made on the basis of spatially and temporally
averaged profiles spatial resolution in the observations is irrelevant.
This is not true for the numerical resolution, which needs to
be sufficiently refined to provide a good match to observed
average line profiles
     \citep{Asplund+Ludwig+Nordlund+Stein2000}.
However, the averaging may mask discrepancies in separate
quantities, like brightness contrast, velocity variations, and size
spectrum of the convective structures that could compensate each other
in the averaging process.

In concert with developments in theoretical modeling, a concomitant
progress in observational techniques of the solar atmosphere has
taken place.
In particular, the advent of high order, high temporal bandwidth
adaptive optics
and post facto image reconstruction techniques, like speckle interferometry
     \citep[e.g.,][]{vonderLuehe1993}
and multi-object multi-frame blind deconvolution
     \citep[e.g.,][]{vanNoort_etal2005},
have greatly advanced our
capabilities of imaging the solar surface with high fidelity.
Together these developments should allow for a promising and more
realistic comparison between theoretical predictions and observations.
Here we attempt to take advantage of developments in instrumentation in
order to compare simulations and observations on the basis
of only the rms contrast, in particular that in the G band.
This eliminates the possible confusion of mixing velocity signatures
with intensity contrast that is inherent in the comparison of spatially
and temporally averaged line profile shapes,
but of course has the drawback that
we have to compare with spatially resolved observations.
Thus we have to compensate for the image degradation imposed by
the Earth's atmosphere,
the telescope, and the instruments on our observed contrast measurements.
To accomplish this we employ a speckle reconstruction technique especially
adapted to provide a realistic reconstruction of image amplitudes, taking
into account the impact of the telescope's Adaptive Optics (AO) system
on the Speckle Transfer Function (STF).

Employing the G band for a comparison between synthetic and observed contrasts
has several advantages.
Imaging in the band is commonly used in high resolution observations
and most observatories nowadays have filters with similar passbands,
so that comparisons can be readily made.
Secondly, because a typical G-band filter is 1 nm wide (FWHM),
it integrates over many lines, masking small uncertainties in individual
line positions and averaging out random uncertainties in line strengths.
The integrated G-band intensity is also insensitive to Doppler shifts
and Zeeman splitting, allowing a comparison of intensity contrast only.
The CH molecular lines that dominate the G band form in the photosphere
where the approximation of Local Thermodynamic Equilibrium (LTE)
applies, which simplifies their numerical modeling.
Moreover, the response of the integrated G-band intensity to temperature
is dominated by the continuum
      \citep{Uitenbroek+Tritschler2006},
to which LTE definitively applies. 

In their overview of realistic solar convection simulations
\citet{Stein+Nordlund2000} compare histograms of granular contrast
observed with the 0.5 m Swedish telescope in the continuum with
the contrast from their simulations, and find very good agreement
after smearing the artificial data with a Point Spread Function
(PSF) representative of the telescope and atmosphere.
\citet{AsensioRamos_etal2006} compare the statistics of spatially
resolved line profiles of the \FeI\ 709.0 nm line observed with
the Interferometric BIdimensional Spectrometer (IBIS) at the DST
with those of profiles computed through a three-dimensional
hydrodynamic snapshot, and find very good agreement for the
intensity distributions in the continuum (after the theoretical
profiles were adjusted for degradation by the telescope and earth
atmosphere) and for the correlation between wavelength dependent
intensity and intensity in the continuum.  However, they also find
that the spread in velocities in the simulation is higher than in
the observed profiles, with the latter lacking velocities beyond
$\pm 1.2$ km s$^{-1}$.  \citet{Keller2006} provides an overview of
the promises and pitfalls of comparing high spatial resolution
observations with magneto-hydrodynamic simulations.

In \S~\ref{sec:observations} we introduce the observational details
and in \S~\ref{sec:speckle} the speckle reconstruction method.  The
simulations are described in \S~\ref{sec:simulations}.  In
\S~\ref{sec:comparison} we compare the theoretical and observed G-band
contrast.  Conclusions are given in \S~\ref{sec:conclusions}.

%
%

\section{Observations\label{sec:observations}}
%
\begin{figure}[tbhp]
  \epsscale{0.5}
  \plotone{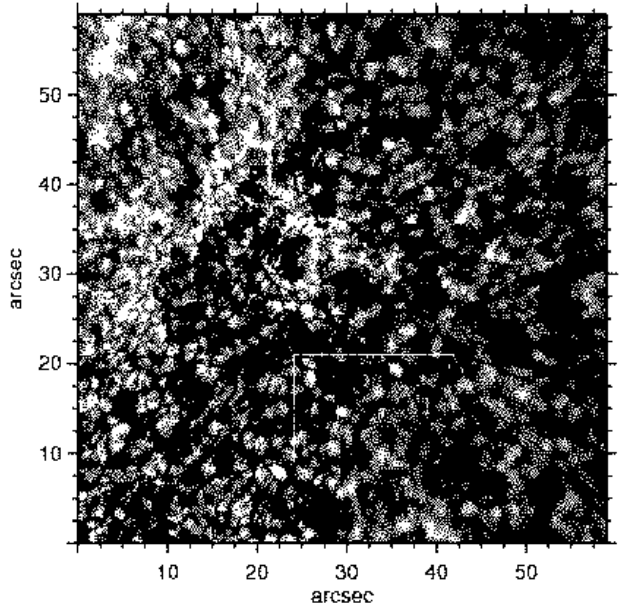}
  \caption{Combination of the reference frame (\emph{left half})
   and burst average (\emph{right half}) of burst No.~9.}
  \label{fig:obs_pore}
  \epsscale{1.0}
\end{figure}
%
The observations were obtained on October 24, 2005 at the Dunn Solar
Telescope (DST) of the National Solar Observatory at Sacramento Peak, Sunspot,
taking advantage of the high-order adaptive
optics system \citep{Rimmele2004} and a high-speed 2k$\times$2k
detector with 12\,$\mu$ pixels manufactured by DALSA for data
acquisition. The optical setup provides a field-of-view (FOV) of about
61\arcsec$\times$61\arcsec. To select the G-band wavelength range a
1.0\,nm wide interference filter centered around 430.5\,nm located in
front of the detector was used. The diffraction limit ($\lambda/D$) of
the DST at this wavelength is 0\farcs117 which together with the
detector image scale of 0\farcs030\,pixel$^{-1}$ leads to an approximate twofold
spatial oversampling of four resolution elements of $\lambda/D$ per pixel,
where $D = 0.76$ m is the aperture of the DST.

\section{Speckle reconstruction\label{sec:speckle}}
In order to allow for speckle reconstruction the data was recorded in
bursts of 80 images, which were acquired within 3.5 sec and streamed off
within 30\,sec. 
The exposure time for an individual frame was 10 ms.
The whole data set comprises 103 bursts with a mean cadence of 67
sec, covering a time period of almost 2\,h in total.
We employ a speckle masking technique, implemented by
     \citep{Woeger2007},
that incorporates an iterative weighted least-squares fitting method for
phase reconstructionand based on a technique by
     \citep{Matson1991}.
In order to get an improved amplitude reconstruction, and thus more
reliable photometry, the reconstruction code takes into account
how the wavefront correction of the AO system affects the STF.

\begin{figure}[tbhp]
  \epsscale{0.6}
  \plotone{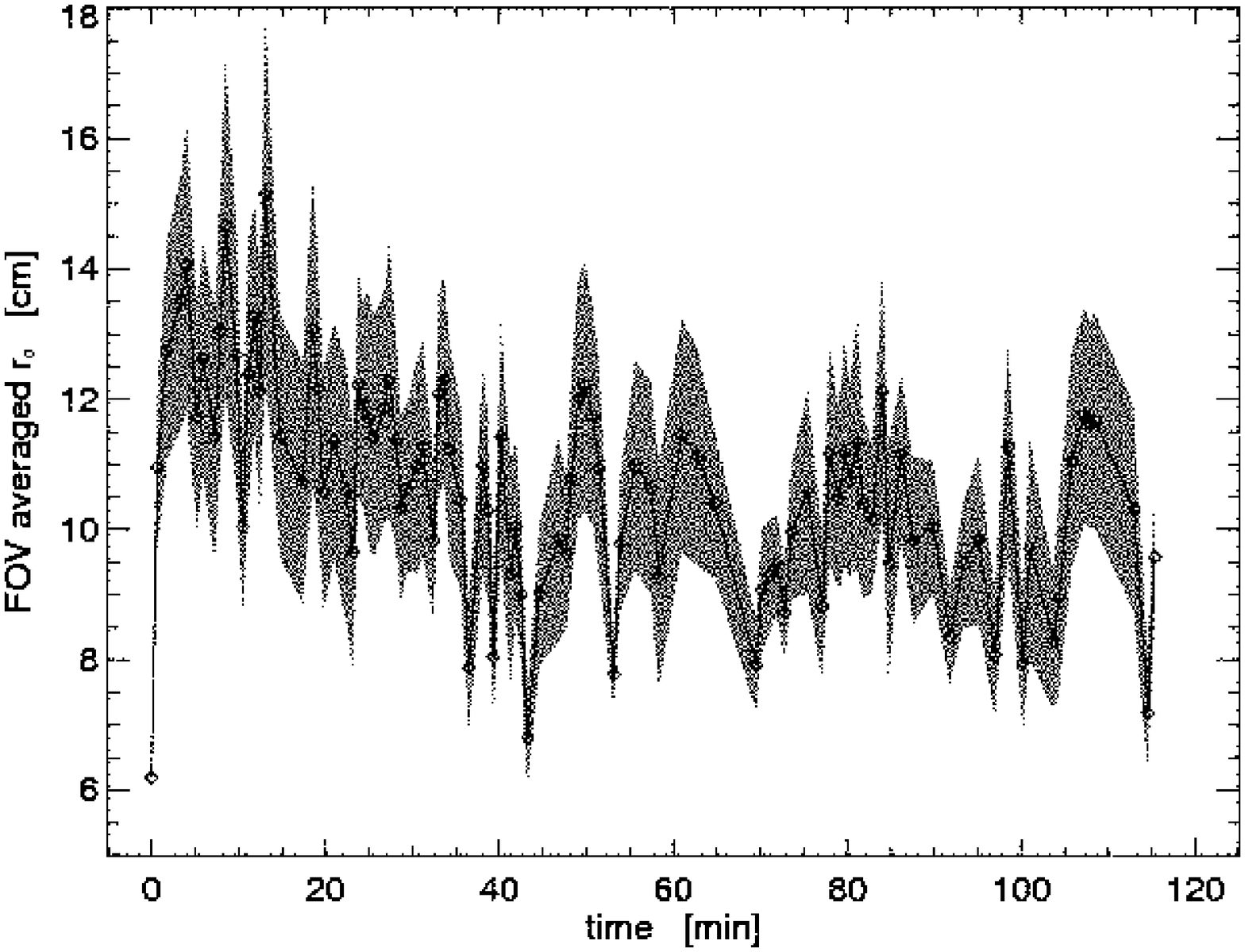}
  \plotone{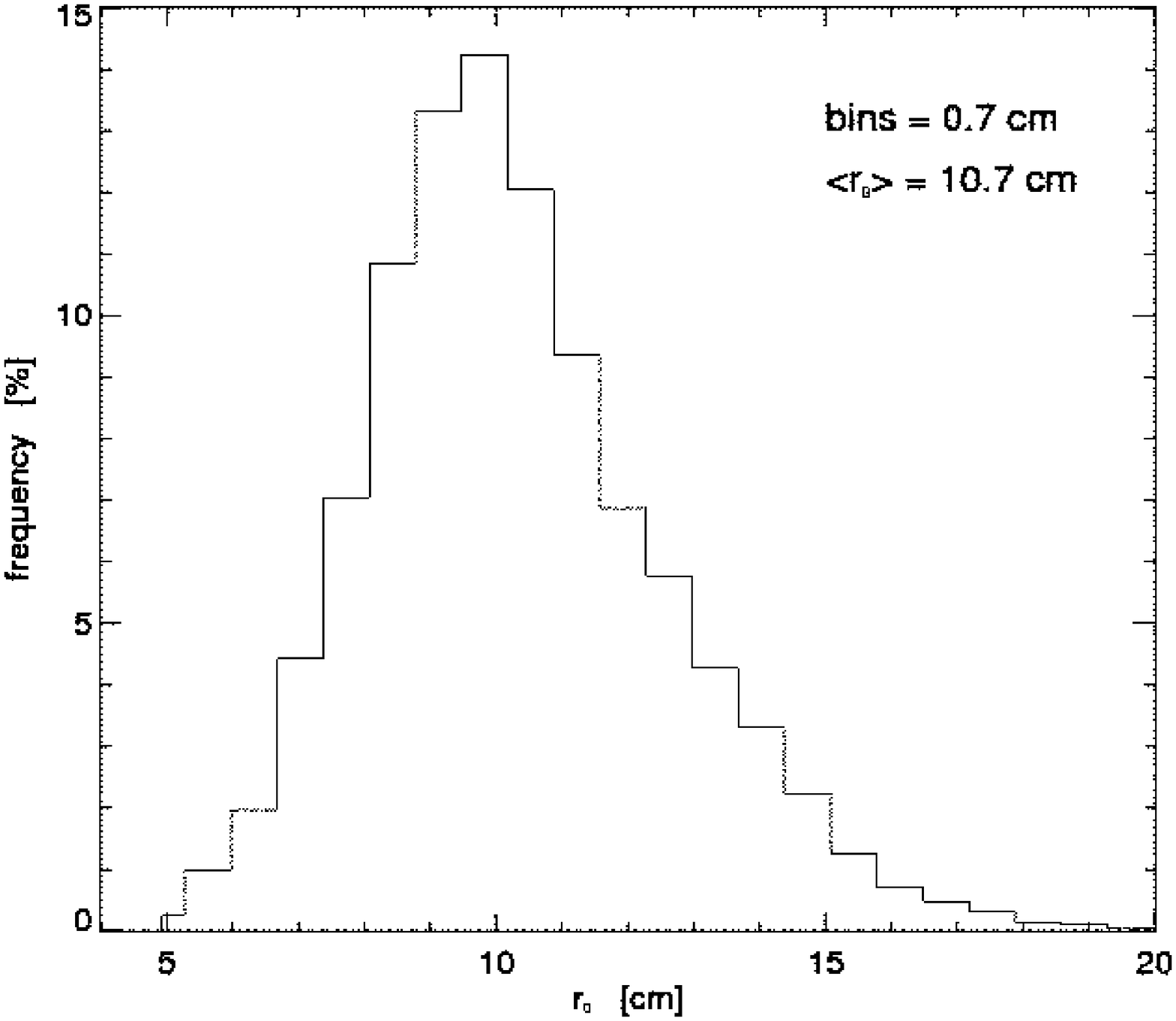}
  \caption{The average Fried parameter $r_0$ over the FOV,
     as determined for each burst
     by the speckle reconstruction code is shown in the \emph{top panel},
     together with its standard deviation (gray shading).
     The \emph{bottom panel} shows the histogram of $r_0$ values determined
     by the speckle code for all $15\times15$ subfields in all bursts.
     \label{fig:subalpha}}
  \epsscale{1.0}
\end{figure}

Each burst frame was divided into 15$\times$15 of 7\farcs68 square each.
The image was then reconstructed on each subfield individually, after which
the corrected subfields were merged back together to give the
complete reconstructed frame.
A byproduct of the speckle reconstruction code is an estimation of the 
Fried parameter $r_0$ over each of the 15$\times$15 subfields in the FOV.
These values characterize the atmospheric seeing over the
duration of each burst with higher values corresponding to more favorable
conditions.
The top panel of Figure~\ref{fig:subalpha} shows the mean value of
$r_0$ for each burst averaged over the subfields
(\emph{solid line and diamonds})
and the standard deviation in the mean (\emph{grey shading}).
A histogram of $r_0$ values per subfield and per burst is shown in the
bottom panel of Figure~\ref{fig:subalpha}.
Both panels make clear that the atmospheric seeing conditions were
good to very good during most of the 2 h sequence.
%

%
%
\begin{figure}[tbhp]
  \epsscale{0.6}
  \plotone{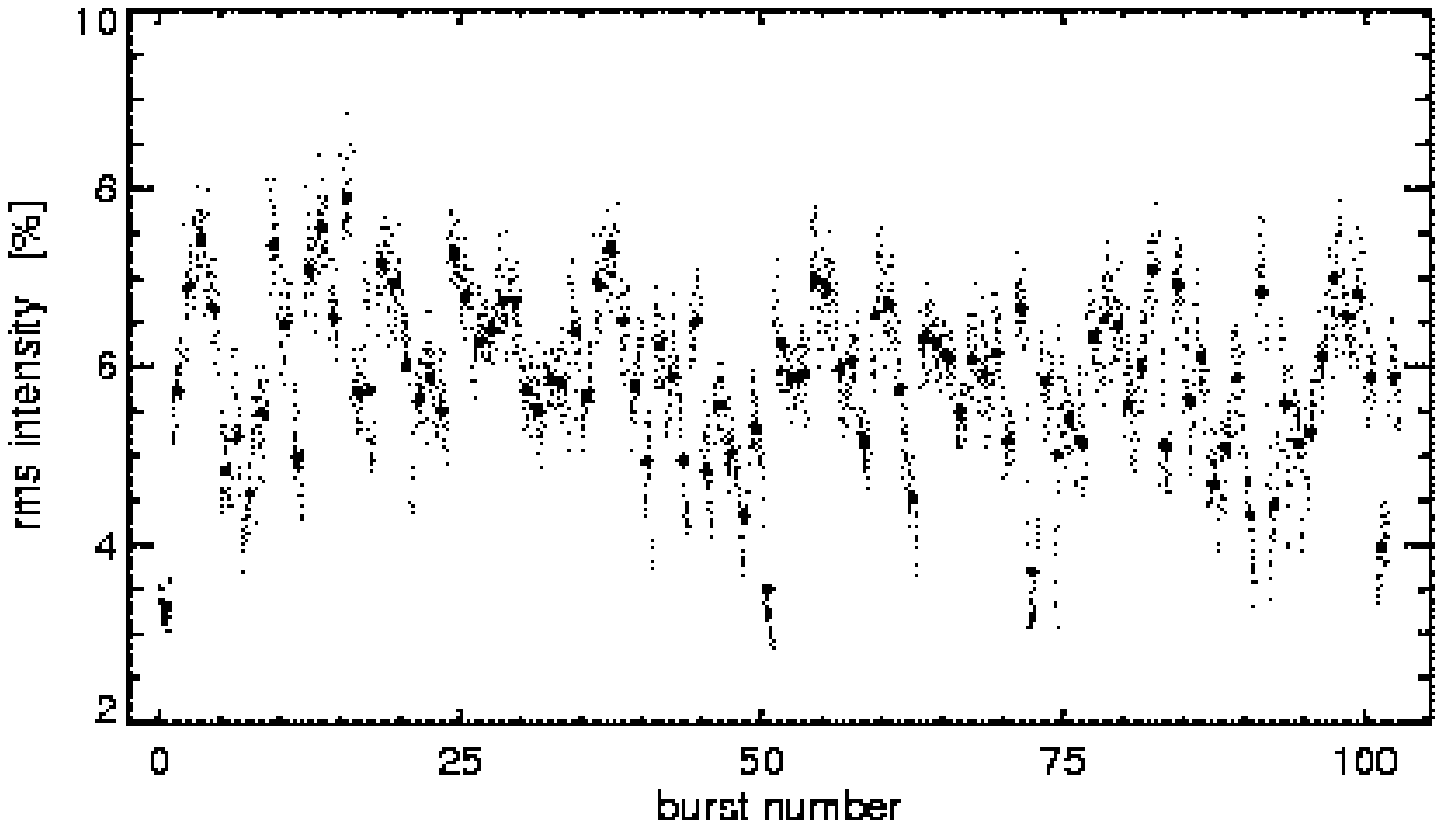}
  \plotone{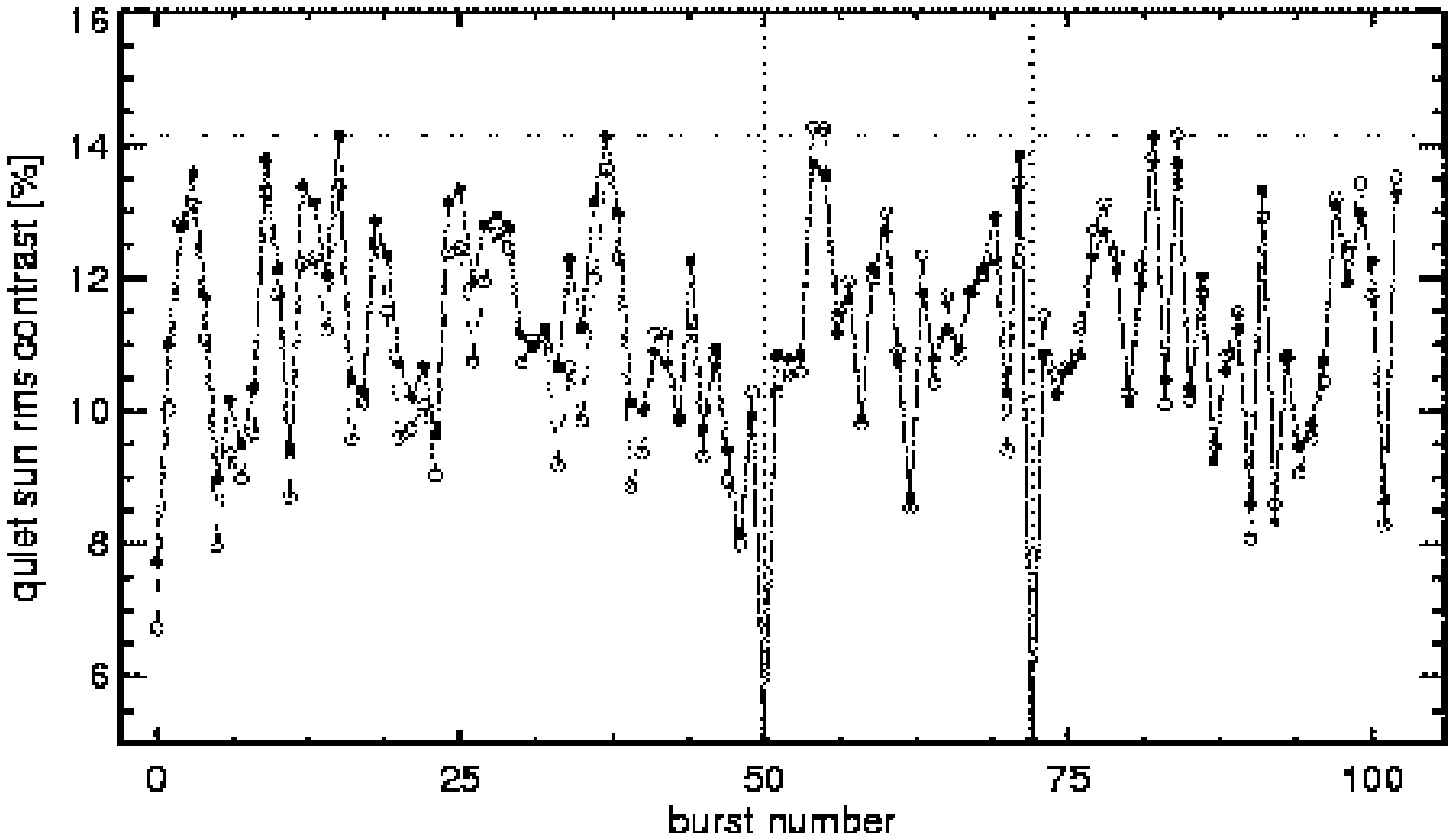}
  \caption{Rms contrast variation in the individual frames for all 103 bursts
    (\emph{top panel}).
     Vertical gray lines indicate the start of each successive burst.
     \emph{Bottom panel} shows the quiet-Sun rms contrast of all
     reconstructed frames, measured from the contrast in a subarea
     (indicated by the white rectangle in Figure~\ref{fig:obs_pore},
     \emph{dashed curve and open circles}), and from the blob algorithm
     (see text, \emph{solid curve and filled circles}).     
     The maximum rms (14.1\%) in the reconstructed images derived
     with the latter is indicated by the horizontal dashed line.
     The vertical dotted lines indicate the two bursts for which
     the iterative reconstruction failed to converge (burst 50 and 72).
     \label{fig:contrasts}}
  \epsscale{1.0}
\end{figure}
%
The top panel of Figure~\ref{fig:contrasts} shows the rms contrast
variation of a quiet area just below the pore (indicated by the white
rectangle in Figure~\ref{fig:obs_pore})
in all 8240 frames of the 103 bursts. The typical uncorrected rms
G-band contrast varies between 5 and 7.5\% in these individual frames,
very rarely reaching values over 8\%.
The selected quiet area is close to the AO lock point, which was
set at the pore, and is relatively devoid of bright points,
and thus is less likely to have strong magnetic fields.
The speckle reconstruction algorithm was successful in 101 of the
bursts. In these successfully reconstructed images the rms contrast 
over the same region of quiet Sun is plotted in the bottom panel
of Figure~\ref{fig:contrasts} (\emph{dashed line with open circles}).
An alternative estimate for the quiet-Sun rms contrast,
more representative of the contrast in the whole frame and not only
close to the lock point, is that provided by a blob finding algorithm
      \citep[see][]{Tritschler+Uitenbroek2006}.
We use this algorithm to locate G-band bright points and eliminate
the pixels encompassed by them from the rms estimate.
The resulting rms values are plotted with the solid curve and filled
circles in the bottom panel of Figure~\ref{fig:contrasts}.
In general the two estimates for the quiet-Sun rms contrast
match each other very well.

Typically, the reconstruction process raises the
rms contrasts in the images by about a factor of two,
to typical values of between 7 and 14\%, with the highest contrast
over the whole image as recovered with the blob algorithm at
values close to this upper bound.
The contrast of the reconstructed images is highly correlated with the
contrast of the burst images.
During "bad seeing" (as judged by the correspondingly low
values of $r_0$ in Figure~\ref{fig:subalpha}, \emph{top panel})
high spatial frequency information is lost and
cannot be recovered by the reconstruction algorithm.
Otherwise, the reconstructed contrast would be consistently close to 14\%. 
The upper limit is reached for a range of "input image
contrasts" of 7-9\%.
This means the reconstructed contrast saturates, which
can only be expected if the spatial power spectrum of the granulation
does not have a high frequency tail at spatial scales below the resolution
limit of the telescope.

\begin{figure}[tbhp]
  \epsscale{0.3}
  \plotone{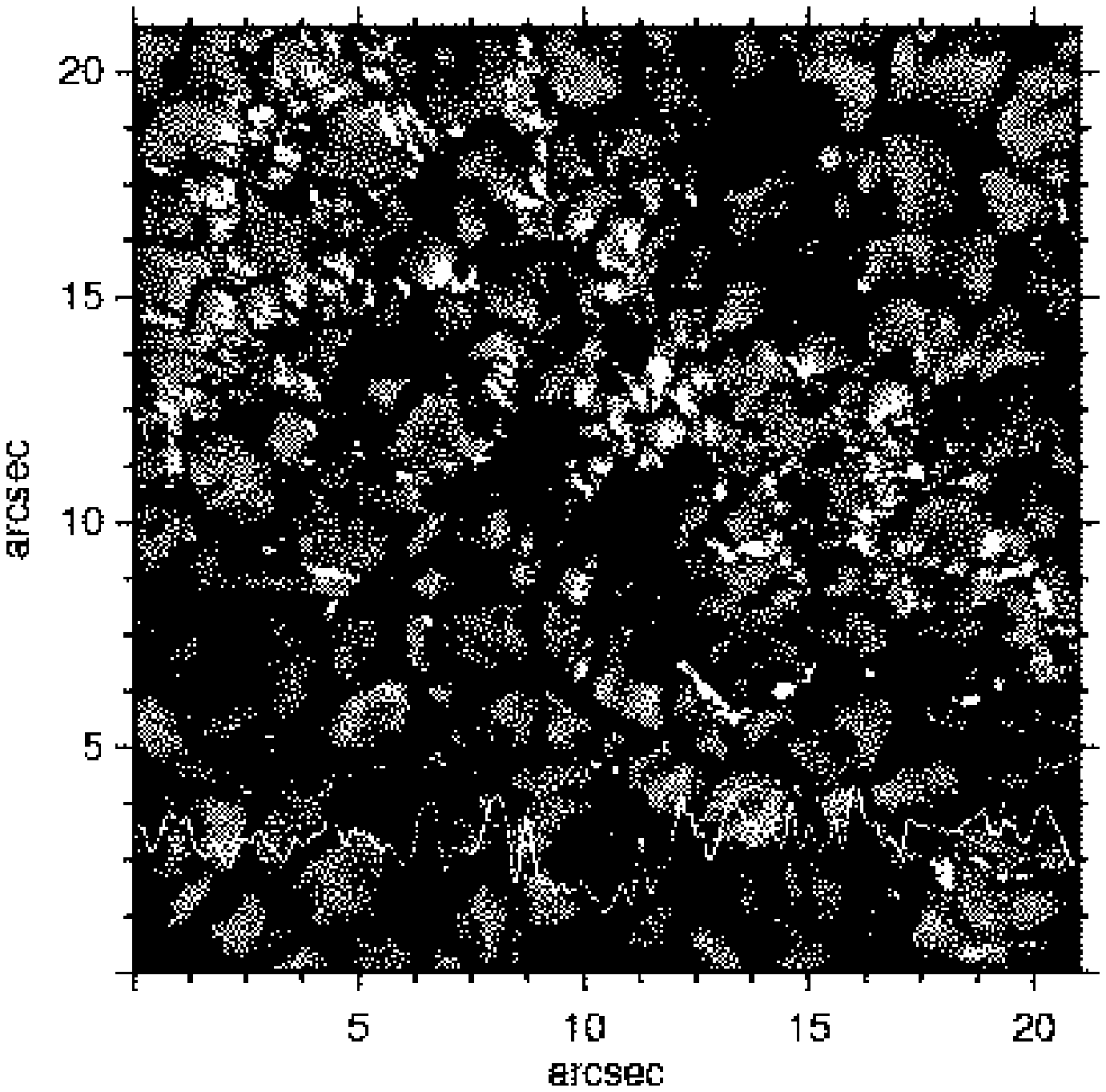}
  \plotone{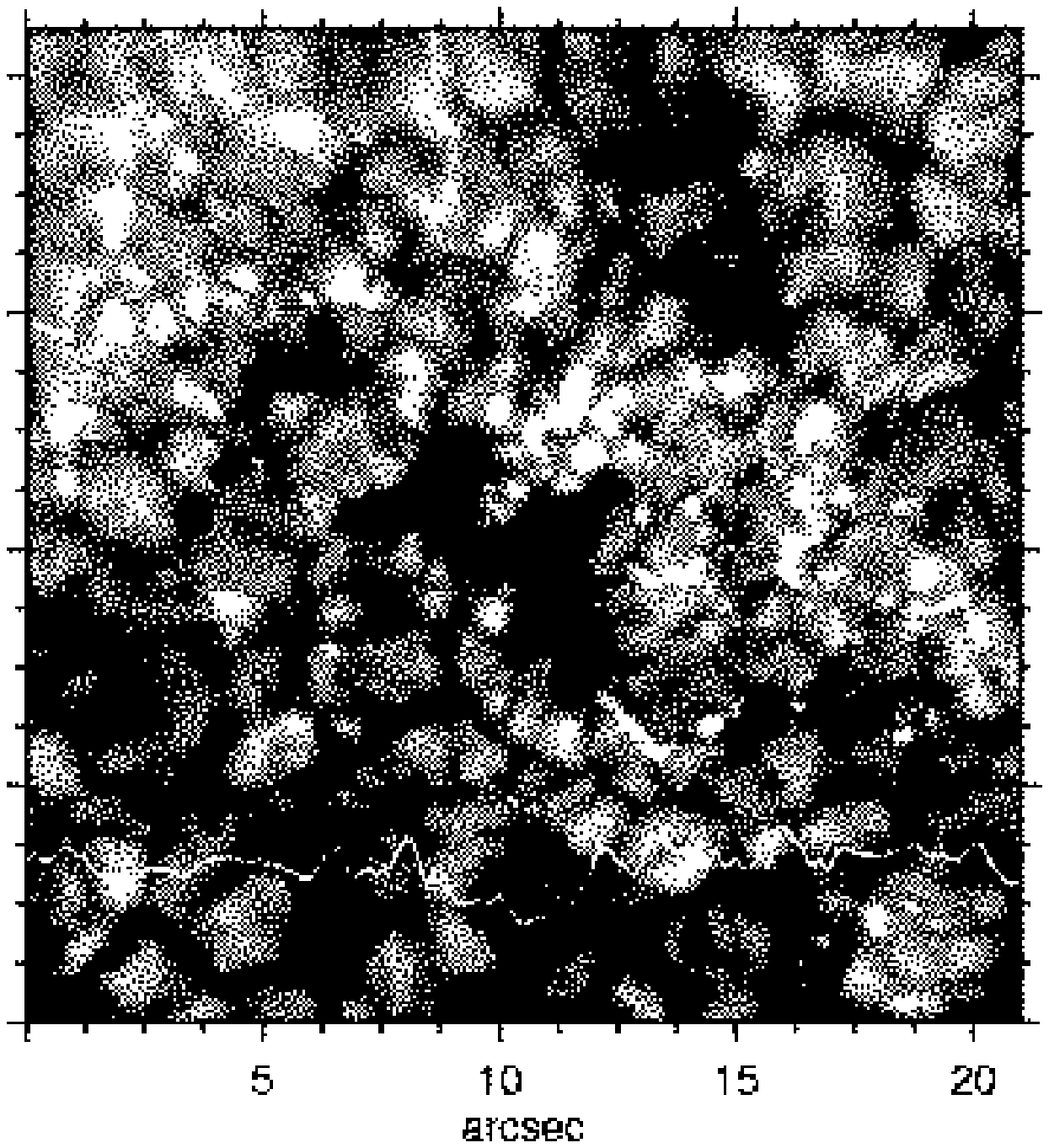}
  \plotone{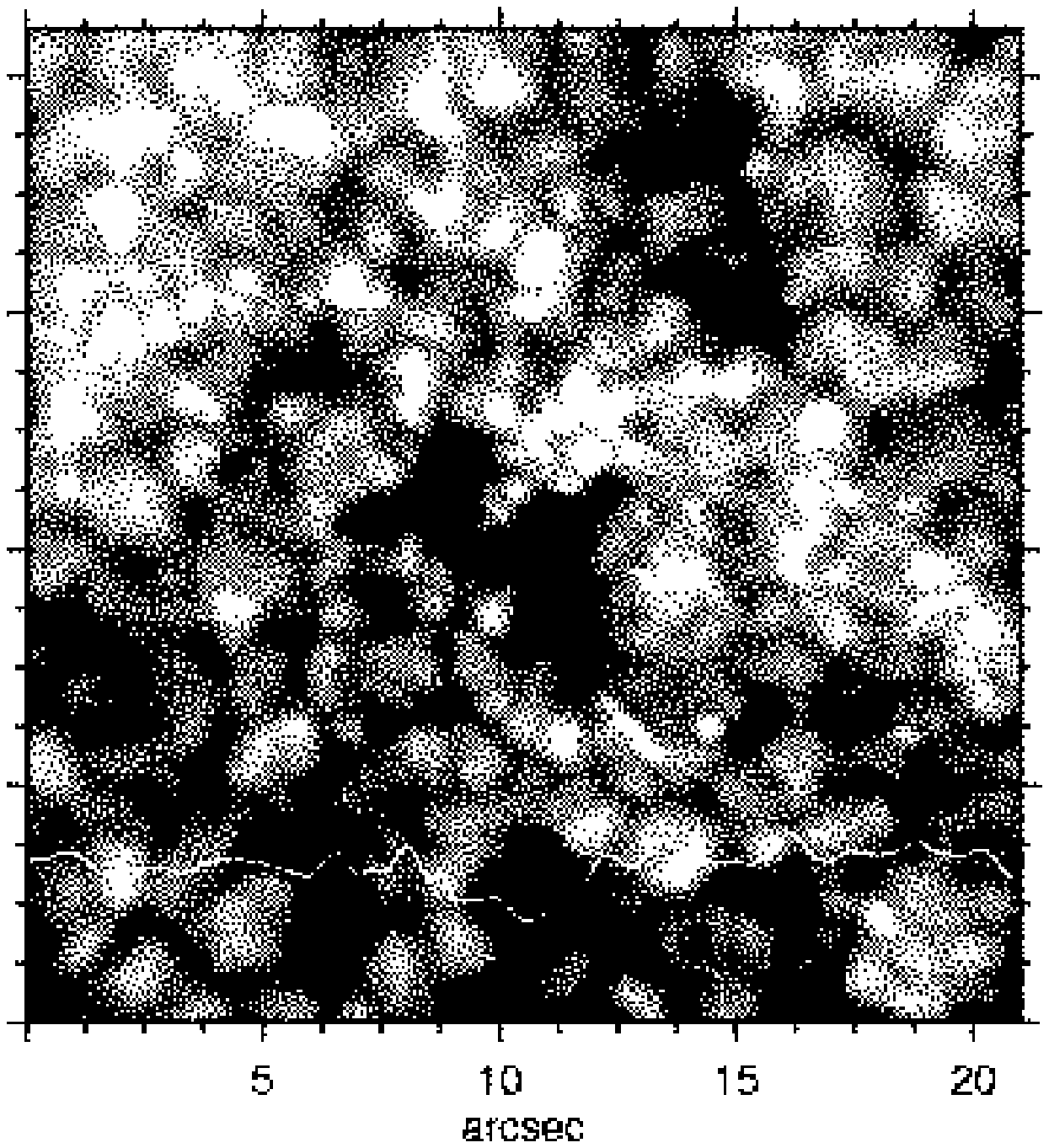}
  \caption{Average (\emph{right panel}), reference (\emph{middle}), and
    reconstructed (\emph{left}) images of the central part of the
    field-of-view centered on the adaptive optics lock point of burst No.\ 15.
    The curve at the bottom of each image represents the G-band intensity
    in a cross section of the image through the pore at 
    $y = 9.9$ arcsec. It illustrates the recovery of the amplitude
    of high spatial frequencies in the reconstructed images.\label{fig:reconstructions}}
  \epsscale{1.0}
\end{figure}

With the doubling of the rms contrast, the improvement
in image quality is clearly illustrated by the sample images
in Figure~\ref{fig:reconstructions}.
The figure shows, from right to left, the central part of a
frame representing the average over the 80 exposures in one burst
(number 15, with the highest average contrast in individual
exposures and in the reconstructed image of the burst),
the reference frame for the reconstruction (the image with the
highest contrast within the burst), and finally the reconstructed image.

%

\section{Simulations\label{sec:simulations}}
We compare the speckle reconstructed G-band images with intensities
obtained from two different simulation snapshots that span a realistic
range in average field strengths on the solar surface outside sunspots.
The first one, here referred to as quiet-Sun has an average 
vertical magnetic field of 30 G, which is probably at the lower limit of
what can be expected in real quiet-Sun areas
      \citep{TrujilloBueno+Shchukina+AsensioRamos2004}.
The second snapshot is representative of solar plage with an average
vertical field of 250 G.
It will be referred to as the plage snapshot.
Both snapshots were taken from magneto-hydrodynamic simulations by
      \citet{Stein+Nordlund2006}.
We employ carbon and oxygen abundances of $\log \epsilon_{C} = 8.39$ and
$\log \epsilon_{O} = 8.66$, respectively as advocated by
     \citet{Asplund+Grevesse+Sauval+AllendePrieto+Blomme2005}
on the basis of \ion{C}{1}, CH, and C$_{2}$ line modeling in
three-dimensional hydrodynamic simulations, and similar
modeling by
      \citet{Asplund+Grevesse+Sauval+AllendePrieto+Kiselman2004}
of \ion{O}{1} and OH lines, and CO lines
      \citet{Scott+Asplund_etal2006}.
The precise value of the oxygen is relevant for the opacity in the G band
region because most carbon in the cool parts of the solar atmosphere is
bound in CO molecules limiting the amount available for CH.

\begin{figure}[tbhp]
  \epsscale{0.75}
  \plottwo{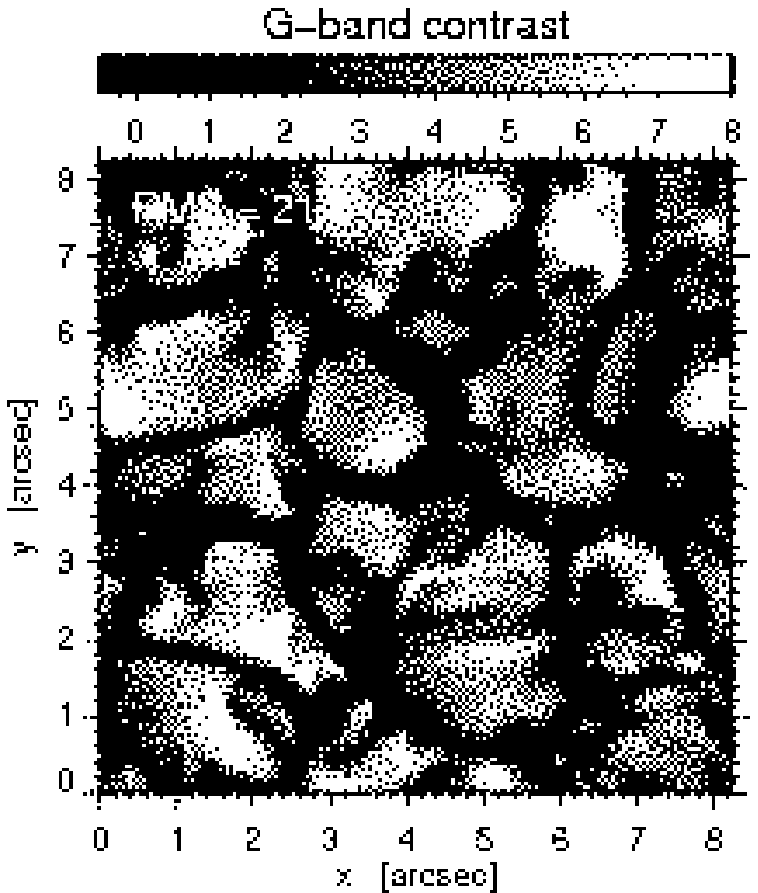}{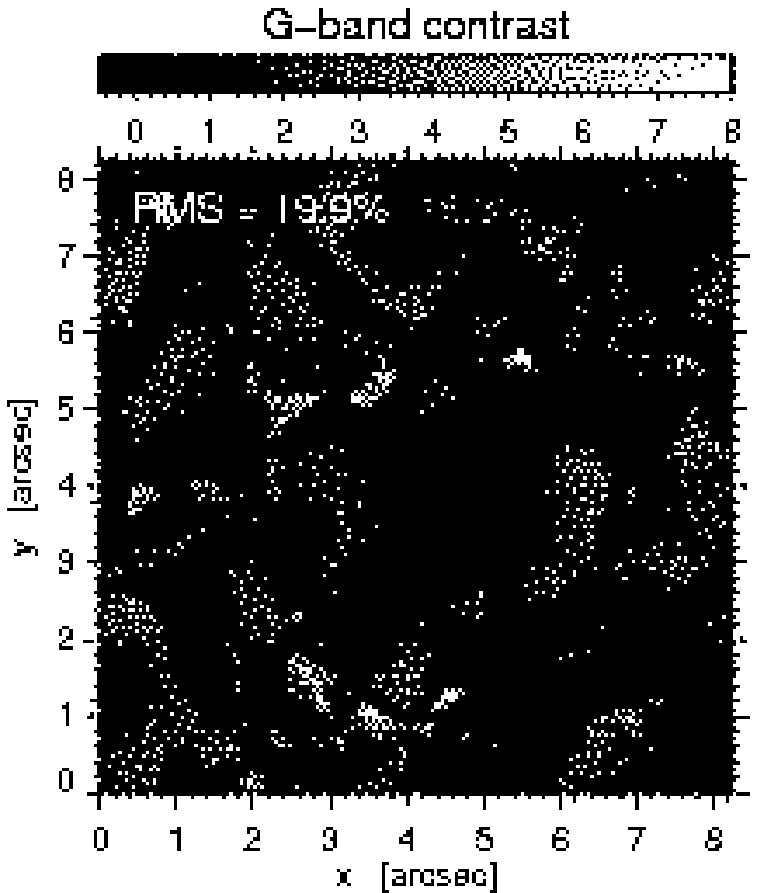}
  \caption{Synthetic G-band filtergrams from quiet-Sun (\emph{left panel})
           and plage (\emph{right}) snapshots.
           Indicated in white are the rms contrast variations over the
           whole filtergrams, and indicated in grey lettering the
           rms contrast of pixels outside bright points in the plage snapshot.
           \label{fig:simulations}}
  \epsscale{1.0}
\end{figure}
%
A description of the spectral synthesis is given in
    \citep{Uitenbroek+Tritschler2006}.
For each of the two snapshots the emergent spectrum in the G-band
region was calculated at 600 wavelengths equidistantly spaced over a range
of 3 nm centered at 430.5 nm.
In the computation this wavelength interval contains 356 atomic lines
and 424 electronic
molecular transitions of the \CHAX\ system with vibrational quantum numbers
$v - v'$ = $0 - 0$, $1 - 1$, and $2 - 2$.
The calculated spectra were then integrated over a typical G-band filter
function of 1 nm FWHM with the shape of a second order generalized Lorentzian,
representative of a dual-cavity interference filter.
The rms intensity differences of the spatially averaged spectrum from the
snapshots with the observed average solar spectrum are only 7\% over the
3 nm spectral range
    \citep{Uitenbroek+Tritschler2006},
confirming the realism of the synthetic spectra.
The main differences between computed and observed spectra result from
slight differences in the central wavelengths of the included spectral lines.
These intensity differences are inconsequential to the wavelength integrated
G-band signal.

Synthetic filtergrams of the two simulation snapshots as well as the
values of their respective rms contrast in the G band are shown in
Figure~\ref{fig:simulations}.
The plage snapshot (\emph{right panel}) includes concentrations of
magnetic field in its inter-granular lanes that are strong enough
to give rise to G-band bright points, while this not the case
anywhere in the quiet-Sun snapshot, despite the presence of
a 30 G average vertical field.
As a result of the bright points the range in contrasts in the plage
snapshot is much higher, and the overall rms contrast is raised as well
     \citep[see also][table 2 therein]{Tritschler+Uitenbroek2006}.
However, the contrast of the plage snapshot measured in the
granulation alone, after eliminating pixels in bright points, is much
reduced compared to the quiet Sun due to the presence of the stronger
magnetic field, which inhibits convective flows, and hence heat
advection.
The granular rms contrast is only 16.3\% compared to 21.5\% in
the quiet-Sun snapshot (see annotations in grey and white,
respectively in Figure~\ref{fig:simulations}),
keeping in mind that the former number is a slight underestimate
because the inter-granular lanes are underrepresented in the rms
through the elimination of bright-point pixels, which occur preferentially
in the dark lanes.

\section{Comparison of observed and theoretical G-band contrasts\label{sec:comparison}}
%
%
\begin{figure}[tbhp]
  \epsscale{0.7}
  \plotone{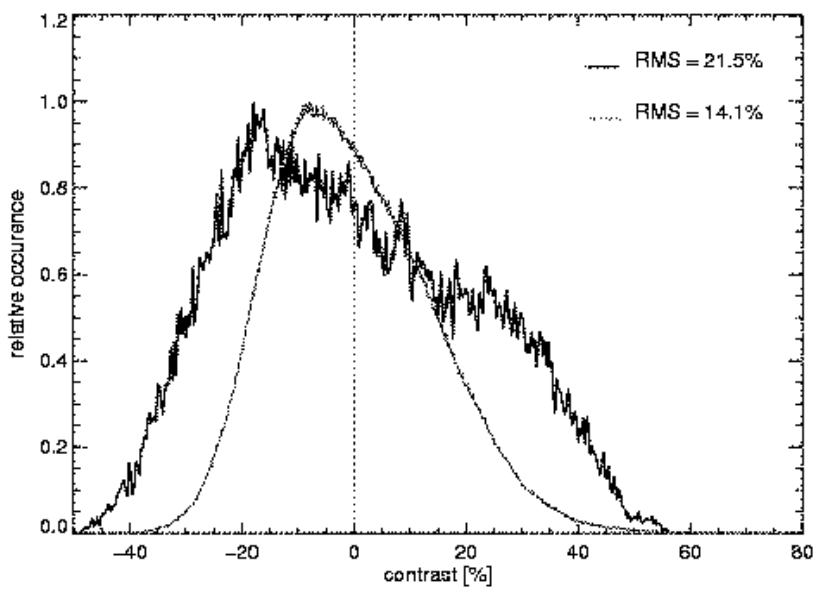}
  \caption{Histograms of G-band contrast values in the
   synthetic quiet-Sun filtergram (\emph{grey curve})
   and best reconstructed image (No.\ 15, \emph{black curve}).\label{fig:histogram}}
  \epsscale{1.0}
\end{figure}
Comparing the rms contrast values between the synthetic
filtergrams and the values for the reconstructed images
(Figure~\ref{fig:contrasts}, \emph{bottom panel})
the large differences are obvious. 
Where the simulations have a granular rms contrasts of 21.5\% for quiet Sun,
the highest contrast in the most quiet regions of the observed reconstructed
G-band images is 14.1\%
This difference is further exemplified in the comparison of
histograms of contrast values in the
synthetic quiet-Sun filtergram (\emph{black curve}) and the reconstructed
frame with the highest contrast (from burst No.\ 15, \emph{gray curve}).
The reconstructed image lacks both relatively dark pixels with contrasts less
than $-30$\%, and relatively bright pixels, with excess contrast of $+30$\%
compared to the synthetic filtergrams.
In addition, the histogram of the synthetic filtergram is more
clearly a bimodal distribution of two Gaussians with a main peak at $-20$\%
and a secondary peak at $+25$\%.
The histogram of the observed filtergram can also be closely represented
by a sum of two Gaussian distributions, but with maxima that are
much closer zero, giving the impression of a unimodal rather
than a bimodal distribution.

%
\begin{figure}[tbhp]
  \epsscale{0.4}
  \plotone{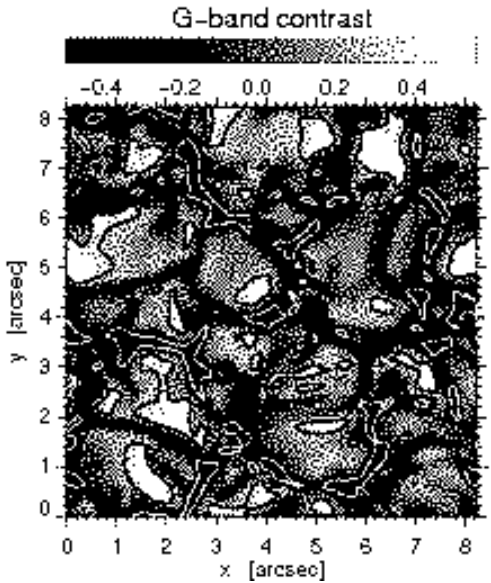}
  \caption{Contours outline the pixels with contrasts
    of less than $-30$\% (\emph{white contours}), and more than $+30$\%
    (\emph{black contours}).\label{fig:darkbright}}
  \epsscale{1.0}
\end{figure}
The darkest (contrast less than $-0.30$, white contours) and brightest
features (contrast less than $+0.30$ black dashed contours) in the
synthetic quiet-Sun snapshot are outlined in Figure \ref{fig:darkbright}.
Clearly, the dimensions of the darkest and brightest areas in particular
are well above the theoretical resolution limit of the DST
(which is $0\farcs117$ at the central wavelength of the G band),
and should be resolvable in the speckle reconstructed images when
seeing is good.
This suggests that the properties of the simulation we
designate as quiet-Sun are inherently different from the real quiet Sun.
In turn this implies that temperature fluctuations in the simulations
are overestimated because granular intensities in the G-band filter
signal are dominated by the continuum and are directly related
to temperature at heights between 25 -- 50 km.
     \citep{Uitenbroek+Tritschler2006}.

\begin{figure}[tbhp]
  \epsscale{0.6}
  \plotone{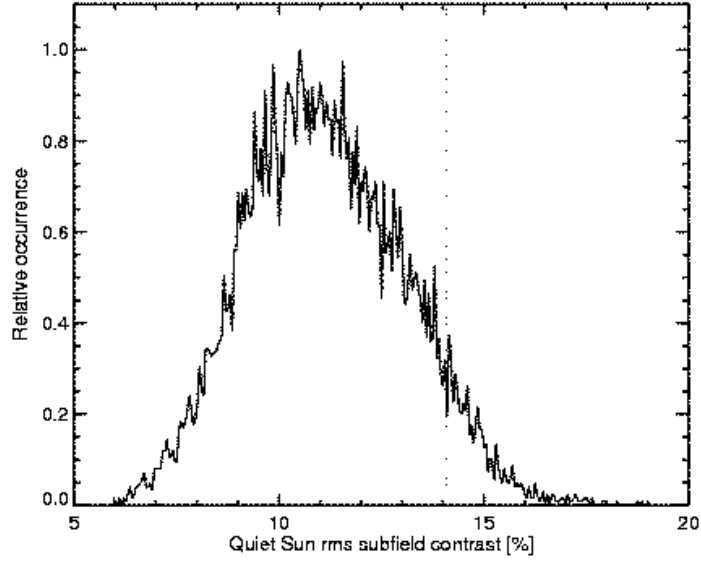}
  \caption{Histogram of rms contrast variation over $8\farcs5\times8\farcs5$
   subfields in the quiet-Sun are of the reconstructed images.\label{fig:subfield}}
  \epsscale{1.0}
\end{figure}

To verify that the quiet-Sun simulation snapshot
does not represent an exceptional case of rms high contrast,
we investigated the statistics of rms contrast of subfields of
comparable size in the observations. 
A histogram of rms contrast values in $8\farcs5\times8\farcs5$
subfields of the 101 successfully speckle-reconstructed images
is shown in Figure~\ref{fig:subfield}.
Only 0.89\% of the subfields have a rms contrast of 14.1\%
(the highest value of rms in the quiet area of the reconstructed frames),
and only 0.01\% has an rms of 20\% or more.
It is therefore highly unlikely that the simulation snapshot we 
employed happens to correspond to the high contrast tail of the
distribution of subfield rms values.
At first sight it seems questionable that less than 1\% of subfields
have rms contrast more than 14.1\%, since there are several frames with
quiet-Sun rms contrast that high (see Figure~\ref{fig:contrasts},
\emph{bottom panel}).
The explanation for this seeming discrepancy is the 5-minute oscillation,
which has patches of intensity variation of typical size 10$''$--15$''$.
Such large patches contribute much more to the overall rms in the frames
than to the rms contrast in subfields smaller than 10$''$.
%

\subsection{Spatially unresolved scattered light\label{sec:scatter}}
One type of image degradation that reconstruction techniques do not have a handle
on is the effect of spatially unresolved scattering, essentially the image
smearing that occurs because of the broad outer wings of the point spread function
(PSF) of the telescope and atmosphere.
Is this perhaps the main reason for the substantially lower contrast of the
reconstructed images with respect to the synthetic filtergrams?
Below we present a simple argument that this is very unlikely the case.

Let $I_{ij}$ the intensity in a synthetic solar image and $\langle I \rangle
= \frac{1}{N} \sum_{ij} I_{ij}$ its average over the $N$ pixels of the image.
The rms contrast $C$ of the intensity is defined as:
\begin{equation}
  C = \sqrt{\frac{\sum_{ij} \left(I_{ij} - \langle I \rangle\right)^2}{%
      \langle I \rangle^2 N}}.
\end{equation}
In the simplest approximation scattering takes a fraction $\alpha$ of the spatially
resolved intensity and spreads it equally over all pixels.
The resulting degraded image $I'$ can therefore be written as $I' = (1 - \alpha) I +
\alpha  \langle I \rangle $, with $\langle I' \rangle = \langle I \rangle$.
The rms contrast of the degraded image is
\begin{equation}
  C' = \sqrt{\frac{\sum_{ij} \left(I'_{ij} - \langle I' \rangle\right)^2}{%
      \langle I' \rangle^2 N}} = (1 - \alpha) C.
\end{equation}
For the calculated contrast to be degraded from the nominal 21.5\% to the observed
contrast in the best reconstructed images of 14.1\% requires a scattering coefficient
$\alpha$ of at least 0.34.
This is unrealistically high for the DST by at least a factor of three, as is judged
from observations off the solar limb, and a transit of the planet Mercury
(Reardon, private communication).

\section{Conclusions\label{sec:conclusions}}
We have compared the rms intensity contrast in the G-band in a series
of speckle reconstructed images obtained at the $D = 0.76$ cm DST
with that in two synthetic G-band filtergrams computed from 
magneto-hydrodynamic snapshots representative of quiet-Sun and plage
with average vertical magnetic fields of 30 G and 250 G, respectively.
We find that the rms G-band contrast in the best speckle-reconstructed
images is just over 14\%, as judged from a quiet subarea, relatively devoid
of bright points, and from the overall contrast in the image after
removing bright points (Figure~\ref{fig:contrasts}, \emph{lower panel}).
This observed rms contrast is contrast is considerably
lower, by more than 50\%, than the rms contrast in the synthetic
quiet-Sun filtergram which was found to be 21.5\%.
In the plage snapshot the rms contrast of the granulation, excluding pixels
with bright points, is 16.3\%.

The employed speckle reconstruction algorithm accounts for the
effect of the telescope's AO system, and provides reliable photometry
in the reconstructions.
During moments of consistently good seeing over the duration
of an 80 frame speckle burst (corresponding to typical values
of the Fried parameter $r_0$ of 12 cm or larger, compare
Figures~\ref{fig:subalpha} and \ref{fig:contrasts}, \emph{top panel}),
the reconstructed rms contrast seems to saturate to 14.1\%,
suggesting this approaches the contrast on the real Sun.
To verify the conclusion that the rms G-band in the observed
quiet Sun is indeed substantially different from the that
in the synthetic snapshot we convolved the latter first
with the telescope Point Spread Function (PSF) appropriate for
the $D = 0.76$ m DST, and then with a PSF representative for
telescope plus $r_0 = 15$ and $r_0 = 10$ cm seeing.
The resulting rms contrasts were 19.3\%, 16.8\%, and 14.9\%,
respectively.
These values are again more than 50\% higher than the rms
contrast in the best quiet-Sun subarea in the uncorrected
single frames (Figure~\ref{fig:contrasts}, \emph{top panel}),
which do not reach more than 9\%, even though they are
estimated close to the AO lock point.
Remarkably, the rms contrast of the quiet-Sun synthetic filtergram,
degraded to the values representative of far from superior seeing
(i.e., $r_0 = 10$ cm), is higher than that in the best reconstructed
images, which should have all image degradation, apart from spatially
unresolved scattering, taken out.

Areas in the synthetic filtergram with negative contrast in excess of
that in the observed quiet Sun are found in the inter-granular lanes,
those with positive excess in relatively large (several tenths of arcsec)
patches in the granules (see Figure~\ref{fig:darkbright}).
The spatial scales of these areas are large enough,
for positive excess in particular, so that they
should be resolved in the best speckle reconstructed images.
The synthetic filtergram, therefore, does not show significant
power that contributes to the rms at spatial scales smaller than
the resolution limit of the DST, which is 0\farcs117 in the G band.
The saturation of the reconstructed rms contrast to just
above 14\% during the moments of best seeing seems to confirm
that no significant power is present in the solar granulation
at spatial frequencies close to the resolution limit of
the telescope.
The simple consideration of degradation of contrast by spatially
unresolved scattered light presented in Section~\ref{sec:scatter}
shows that it is very unlikely the cause of the discrepancy
between computed and observed G-band contrasts.
We remark that, given the typical spatial scale of intensity fluctuations
caused by the 5 min oscillations, the latter contribute more
to the rms of the large FOV spanned by our images than to the
rms contrast in small fields like our simulation snapshots,
which are smaller than those $p$-mode spatial scales.
Therefore, the discrepancy in G-band contrast caused by convective
motions alone is even larger than it seems from our estimates.
We conclude that there remains a clear and unresolved difference
between the simulated and observed rms G-band contrasts:
the quiet-Sun simulation snapshot overestimates the rms
G-band contrast by as much as 50\%.
This raises the fundamental question whether the simulations
still lack fundamental physics or sufficient numerical resolution,
or whether we do not fully understand the degradation of contrast
by the earth atmosphere and telescope during observations.

One possible source of the contrast discrepancy is the absence
of cooling by CO molecules in the simulation, where radiative
transfer is approximated by only four opacity bins, and four
directions.
CO is the dominant coolant above the photospheres of
cool stars 
      \citet[e.g.,][]{Ayres+Testerman1981},
and its presence may well reduce temperature
differences in that layer, leading to a reduction in rms
contrast variation in the G band since the latter is
directly related to temperature variations at about 25-50 km
above the photosphere.
However, the temperature reduction by CO may only be important
in layers that are higher than that
      \citep{Scott+Asplund_etal2006}.

The clearest hint, however, towards a solution to the
G-band contrast discrepancy comes from the much reduced granular
contrast in the plage simulation snapshot.
Clearly, the presence of high concentrations of magnetic field
inhibits convective motions and convective heat transport,
and reduces temperature differences and thereby, G-band contrast.
The findings described by
    \citet{AsensioRamos_etal2006}
that simulated profiles of the \FeI\ 709.0 nm line in a 
purely hydrodynamic simulation
     \citet{Asplund+Ludwig+Nordlund+Stein2000}
show high velocties in excess of $\pm 1.2$ km s$^{-1}$ that are lacking
in high resolution observed quiet-Sun observations confirms that
such simulations without magnetic field seem to overestimate
convective flow velocities.
We speculate that the real quiet Sun contains much more magnetic field
than the average 30 G that we assumed in our ``quiet-Sun'' snapshot,
so that the comparison of its rms G-band contrast to that in
the observed quiet-Sun is not appropriate, much like comparing
apples to oranges.
This magnetic field has to be present mainly in the form of weak fields,
since it does not show up in the form of bright point in the G band.
We further speculate that, if indeed the real Sun contains more
magnetic field and if convective flow speeds are less than
what is derived from field-free simulations, the element abundances
that are derived from such simulations are underestimated.
For slower convective flows abundances would have to be increased
to match the observed widths of lines.
This would alleviate, at least in part, the conflict that exists
between the low abundances of oxygen and carbon
     \citep{Asplund+Grevesse+Sauval+AllendePrieto+Blomme2005,%
Asplund+Grevesse+Sauval+AllendePrieto+Kiselman2004,Scott+Asplund_etal2006},
derived from three-dimensional simulations compared to their
meteoritic values, and the conflict these low values
pose with current models of the solar interior
     \citep[eg., ][]{Ayres+Keller+Plymate2006}.

For more detailed comparison between observed and simulated G-band
contrast more simulations are required with intermediate field
strengths between the 30 G and 250 G average values we employed here.
In addition, observations from space, where the degradation of image
quality by the earth atmosphere is eliminated, and only the easier
to characterize contrast reduction caused by the telescope and instrument
has to be considered, should conclusively establish the value
of the G-band contrast of the real Sun, hopefully confirming
our conclusions here.
This has to be accompanied by deep measurements of the magnetic
(vector) field, so that we can finally answer the question what
really constitutes quiet Sun.

\acknowledgements We are grateful to Bob Stein for providing
the simulation snapshots, and to Friedrich W\"oger for allowing
us to use his speckle reconstruction code.

\end{document}